\documentclass[10pt,aps,prb,twocolumn,groupedaddress,showpacs]{revtex4-1}
\pdfoutput=1
\usepackage{amssymb,amsmath}
\usepackage{graphicx}
\usepackage{color}
\usepackage{hyperref}

\newcommand{\digammaa}{\psi}
\newcommand{\trigamma}{\psi'}

\newcommand{\tphi}{\ensuremath{\tau_\phi}}
\newcommand{\tphii}{\ensuremath{\tau_\phi^{-1}}}

\newcommand{\lcut}{\ensuremath{B_{\rm sm}}}
\newcommand{\kT}{\ensuremath{k_B T}}
\newcommand{\ltotal}{\ensuremath{B_{\rm tot}}}

\DeclareMathOperator{\sech}{sech}

\begin{document}
\title{Conductance fluctuations in quasi-two-dimensional systems: a practical view}


\author{M. B. Lundeberg, J. Renard, and J. A. Folk}
\affiliation{Department of Physics and Astronomy, University of British Columbia, Vancouver, BC, V6T 1Z1 Canada}

\date{\today}
\begin{abstract}
The universal conductance fluctuations  of quasi-two-dimensional systems are analyzed with experimental considerations in mind.
The traditional statistical metrics of these fluctuations (such as variance) are shown to have large statistical errors in such systems.
An alternative characteristic is identified, the inflection point of the correlation function in magnetic field, which is shown to be significantly more useful as an experimental metric and to give a more robust measure of phase coherence.
\end{abstract}


\maketitle

Universal conductance fluctuations (UCF) are the static but sample-dependent variations in electrical conductance that arise in diffusive mesoscopic systems due to quantum interference.\cite{lsf,bergmann,mpep} Certain aspects of UCF are highly sensitive to phase coherence, and so the analysis of UCF is an important experimental tool in determining the decoherence resulting from a dynamic environment, which is a key factor in any quantum device. Although there are other transport signatures of coherence, they either require a specific sample geometry (e.g., Aharonov-Bohm rings\cite{birgeAB}) or  fail to provide coherence information in environments lacking time-reversal symmetry (this is the case with weak localization, for example\cite{ripples}).
The statistics of UCF, on the other hand, provide a general-purpose measure of coherence for any mesoscopic diffusive device.

In the past, experimental studies of UCF have primarily been carried out with quasi-one-dimensional (quasi-1D) samples, i.e.~wires or channels that are very narrow compared to the coherence length.
The theory of UCF is well developed for this case.\cite{beenakker1d,chandrasekhar} Unfortunately, the quasi-1D regime has several practical drawbacks that have limited its applicability as a tool to study coherence. First, the typical scale of conductance fluctuations in magnetic field is large (hundreds of millitesla in some cases) and hence very large field ranges are required for a proper statistical analysis.\cite{mohantyUCF} In semiconducting systems, the useful field range is often limited by the onset of Landau quantization, so it may be impossible to gather sufficient statistics for a reliable measurement.\cite{taylor} The requirement of large fields also obscures features of the coherence time that may be field dependent.\cite{mohantyUCF}
Another issue is that the system must be very narrow to avoid a complicated crossover to the quasi-two-dimensional (quasi-2D) regime, so edge effects can become important.

The quasi-2D regime, in which the coherence length is smaller than the device lateral dimensions, does not share the disadvantages of the quasi-1D regime mentioned above. Although the fundamental theory of UCF has been well-established for over 20 years,\cite{lsf} to this day its consequences for the quasi-2D case are only partially resolved.\cite{bergmann}

The goal of this manuscript is to present a detailed analysis of the quasi-2D UCF correlation function, with an eye toward using this statistical analysis in experiments.  It is shown that phase coherence information can be reliably extracted only from the correlation function with respect to magnetic field.  
The traditional metrics of this correlation function---variance and half-width---are shown to have unexpectedly large statistical errors, which limit their utility in an experiment.
The inflection point of the magnetic field correlation function, on the other hand, turns out to be a particularly robust measure of coherence, which extends the practical use of UCF as a coherence detector to quasi-2D systems.
The applicability of UCF as a thermometer\cite{falkoucf} in the quasi-2D case is also discussed.

Section \ref{sec:single}  identifies several experimentally-accessible metrics of the quasi-2D UCF correlation function and describes their dependence on decoherence and temperature.
Section \ref{sec:errors} investigates the statistical errors that naturally arise in the analysis of UCF data, and how those errors affect the accuracy of various techniques for measuring coherence from UCF correlations.
Section \ref{sec:symmetry} discusses the adjustment of the correlation function that occurs when there are static symmetry-breaking interactions, such as spin-orbit interaction.
Section \ref{sec:1d} compares these results to the quasi-1D case.

\section{\label{sec:single}Correlation lengths of quasi-2D conductance fluctuations}

A typical UCF experiment begins with measuring conductance $G(\mu,B)$, which is a function of the externally controlled chemical potential $\mu$ and the magnetic field $B$ perpendicular to the sample plane.
The ensemble-averaged conductance background $\overline{G}$ is subtracted off to yield the conductance fluctuations $\delta G(\mu,B) = G(\mu,B)- \overline{ G(\mu,B) }$. The fluctuations contain a large amount of information, but most of it is sample specific and chaotically sensitive to the exact disorder configuration.
The statistics of $\delta G$, on the other hand, are not sample specific and are encapsulated by its correlation function,\cite{lsf}
\begin{equation}
F(\delta \mu, \delta B) \equiv \overline{ \delta G(\mu,B)\, \delta G(\mu+\delta \mu, B + \delta B)}.
\label{ucfcorrdef}
\end{equation}
which is independent of $\mu$ and $B$ under usual conditions, but does depend on temperature $T$, dephasing rate $\tphii$, and diffusion constant $D$.\footnote{The magnetic field should be large enough to break time reversal symmetry, $B \gg \hbar\tphi^{-1}/eD$, yet small enough to not introduce high-field (Landau quantization) effects. The device properties ($\tphi$, $D$, etc.) should not change significantly over the measured range of $\mu$, $B$.}

The theoretical two-parameter correlation function can be computed using various methods (see Refs.~\onlinecite{lsf},\onlinecite{bergmann}, and Appx.~\ref{sec:2dtheory}), and in principle can be compared directly to its experimental counterpart in order to extract temperature or phase coherence information.
In practice, however, experimentalists usually analyze one-parameter cross sections for convenience: $F_\mu(\delta \mu) = F(\delta \mu, 0)$, or $F_B(\delta B) = F( 0, \delta B)$.
We first focus on the magnetic field cross section, $F_B(\delta B)$.

\begin{figure}[t]
\includegraphics{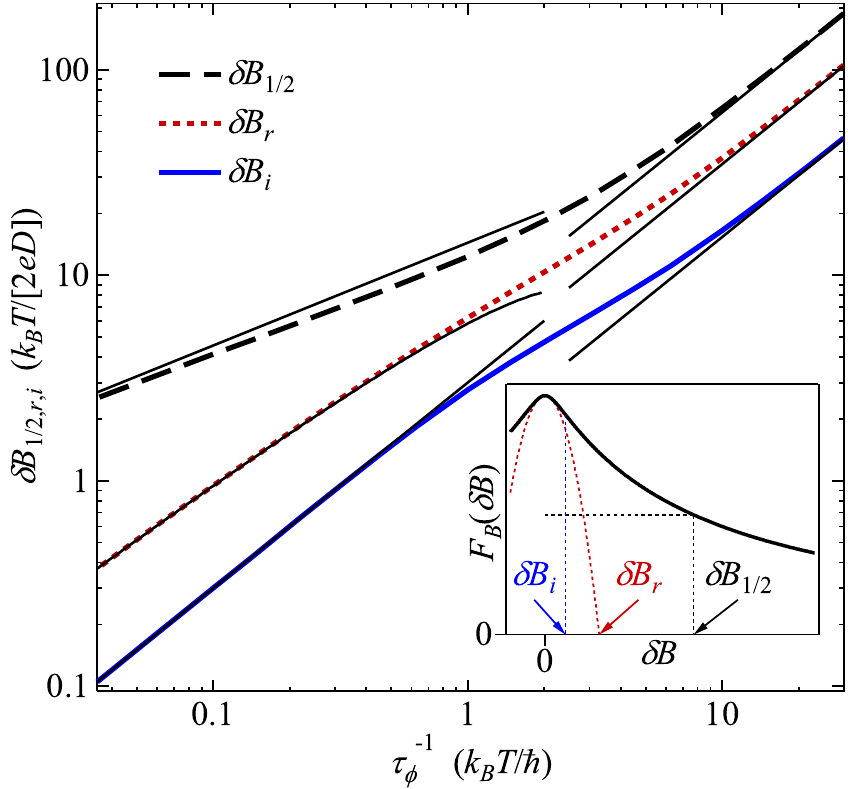}%
\caption{\label{fig-bmap}Dephasing rate dependence of several characteristic scales of the quasi-2D UCF correlation function in magnetic field. Results for inflection point $\delta B_i$ (solid blue), roundness $\delta B_r$ (dotted red), and half-width $\delta B_{\frac{1}{2}}$ (dashed black) are indicated by thick lines. Thin lines indicate the asymptotic forms in Table~\ref{table-asymptotics}. The inset shows the correlation function for $\kT = 3\hbar\tphi^{-1}$ and graphically depicts the definitions of $\delta B_{\frac{1}{2}}$, $\delta B_r$, and $\delta B_i$.}
\end{figure}

It is customary in UCF studies to characterize $F_B(\delta B)$ by its half-width, and compare to values provided by theory.  In the quasi-2D case, however, different field scales associated with the correlation function depend in different ways on $\tphii$, and it is therefore important to identify which scale is most appropriate for a particular experiment.
We consider three different field scales of the correlation function (Fig.~1 inset):
\begin{itemize}
\item The half-width $\delta B_{\frac{1}{2}}$, defined by $F_B(\delta B_\frac{1}{2}) = \frac{1}{2} F_B(0)$, is the point where correlation has fallen to 50\% of the variance.
\item The roundness $\delta B_r = |2F_B(0) / F_B''(0)|^\frac{1}{2}$ characterizes correlations at very small field separation, where $F_B''(\delta B)=\mathrm d^2 F_B/\mathrm d\delta B^2 $.
\item The inflection point $\delta B_{i}$, defined as the point where $F_B''(\delta B_i) = 0$, is the field separation at which correlation falls the fastest (the minimum $F_B'(\delta B)$).
\end{itemize}

Figure \ref{fig-bmap} shows how these three field scales, calculated from the theoretical $F_B(\delta B)$, depend on $\tphii$ in the case that only one dephasing rate is relevant (the case of multiple dephasing rates is discussed in section \ref{sec:symmetry}). 
Immediately one can see that the three scales are not proportional, illustrating the multi-scale nature of $F_B(\delta B)$.   
The field scales are expressed here in terms of a characteristic thermal field $\kT/(2eD)$, and the dephasing rate in terms of the thermal time,  to make this plot general for any quasi-2D system.

 \begin{table}
 \caption{\label{table-asymptotics}Asymptotic field and energy correlation lengths. Prefactors have been numerically determined, whereas exponents are analytically derived (see Appx.~\ref{sec:longrange}).}
 \begin{ruledtabular}
 \begin{tabular}{rcc}
  & Smeared limit & Unsmeared limit  \\
Measure & $\kT \gg \hbar \tphii$ &  $\kT \ll \hbar \tphii$ \\
 \hline
 $\delta B_{\frac{1}{2}}\cdot 2eD$ & $14.4(\kT\hbar\tphi^{-1})^{\frac{1}{2}}$ & $6.21 \hbar\tphi^{-1}$ \\
 $\delta B_r\cdot 2eD$ & $\big(24 \ln \frac{4.1 kT\tphi}{\hbar}\big)^{\frac{1}{2}}\hbar\tphi^{-1}$ & $3.48 \hbar\tphi^{-1}$ \\
 $\delta B_i\cdot 2eD$ & $3.01 \hbar\tphi^{-1}$ & $1.53 \hbar\tphi^{-1}$ \\
 $\delta \mu_{\frac{1}{2}}$ & $2.72 \kT$ & $1.67 \hbar\tphi^{-1}$ \\
 $\delta \mu_r$ & $ 3.16 \kT$ & $1.36 \hbar\tphi^{-1}$ \\
 $\delta \mu_i$ & $ 2.14 \kT$ & $0.68 \hbar\tphi^{-1}$ \\
 \end{tabular}
 \end{ruledtabular}
 \end{table}

Table \ref{table-asymptotics} lists the asymptotic behaviour of each field scale in the thermally smeared limit ($\hbar\tphii \ll k_B T $) and the unsmeared limit ($\hbar\tphii \gg k_B T $). The unsmeared limit is rarely encountered at low temperatures, where dephasing is typically dominated by the contribution of electron-electron interactions\cite{eeucf} giving $\tphii = \alpha \kT/\hbar$, for some $\alpha$ less than unity. In the smeared limit, $\delta B_\frac{1}{2}$ depends  equally on $T$ and $\tphi$ (cf.~Ref.~\onlinecite{bergmann}), while roundness $\delta B_r$ depends logarithmically on $T$. Remarkably, $\delta B_i$ has no direct $T$-dependence in either limit. This is a desirable characteristic because a measurement of $\delta B_i$ then yields the value of $\tphii$ directly, without needing exact knowledge of $T$.

\begin{figure}[t]
\includegraphics{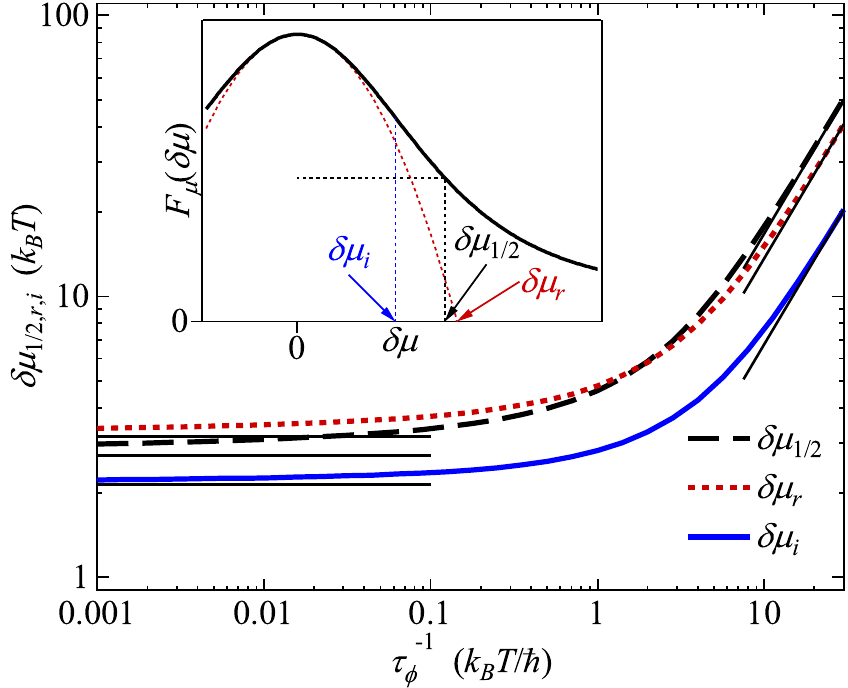}
\caption{\label{fig-emap}Mapping of several energy correlation lengths to dephasing rate, analogous to Fig.~\ref{fig-bmap}. Results for $\delta \mu_i$ (solid blue), $\delta \mu_r$ (dotted red), and $\delta \mu_{\frac{1}{2}}$ (dashed black) are indicated by thick lines. Thin lines indicate the asymptotic forms in Table~\ref{table-asymptotics}. The inset shows the correlation function for $\kT = 3\hbar\tphi^{-1}$ and graphically depicts the definitions of $\delta \mu_{\frac{1}{2}}$, $\delta \mu_r$, and $\delta \mu_i$.}
\end{figure}

Energy correlation lengths $\{\delta \mu_{\frac{1}{2}},\delta \mu_r,\delta \mu_i\}$ are shown in Fig.~\ref{fig-emap}, computed from the theoretical $F_\mu(\delta \mu)$, following definitions analogous to the $\delta B$ correlations. Asymptotic forms are listed in Table \ref{table-asymptotics}.
For strong thermal smearing  ($\kT \gg \hbar\tphii$), $F_\mu(\delta \mu)/F_\mu(0)$ approaches a universal function of $\delta \mu/(\kT)$, which is independent of $\tphii$; this gives the universal correlation lengths $\delta \mu_{\frac{1}{2},r,i}$ listed in Table \ref{table-asymptotics}.  As a result, $F_\mu(\delta \mu)$ is not useful for measuring $\tphii$, but can instead be used as a thermometer.\cite{falkoucf}

The strategy of using UCF as a primary thermometer has been shown to be effective for quasi-1D systems.\cite{falkoucf} For the quasi-2D correlation function, however, convergence to the universal form is very gradual, and all three metrics deviate significantly from their asymptotic values even when $\hbar\tphii < 0.1 k_B T$ (Fig.~\ref{fig-emap}).  The deviation is particularly severe for the metric $\delta \mu_\frac{1}{2}$ identified in Ref.~\onlinecite{falkoucf},  e.g.,~20\% at $\hbar\tphii = 0.05 k_B T$.  Somewhat faster convergence is observed for the inflection point $\delta \mu_i$ (e.g.,~8\% deviation at $\hbar\tphii = 0.05 k_B T$). For the highest accuracy, both $\delta B_i$ and $\delta \mu_i$ should be measured: together these provide unique values for both $T$ and $\tphi$.

\section{\label{sec:errors}Statistical errors in quasi-2D UCF measurements}

Statistical errors play a major role in experimental studies of UCF.
Even when the conductance $G(\mu,B)$ is measured exactly (without noise), there are two statistical error sources that affect the quality of the analysis: random errors due to a limited data set, and systematic errors due to the background subtraction procedure.
A formal treatment of these errors can be found in Appendix~\ref{sec:errorstheory}.
 
In this section we examine the two types of errors as they apply to experiments measuring $F_B(\delta B)$ from quasi-2D UCF in magnetic field.  It is worth noting that errors associated with measurements of $F_\mu(\delta \mu)$ are very different, but this function contains little phase coherence information.
A careful statistical analysis is required to estimate the errors in $F_B(\delta B)$ with any degree of accuracy---a nonintuitive result that comes from the multi-range nature of the function. 
Inspection alone generally overestimates the number of effectively independent samples that contribute to an averaged correlation function, often by orders of magnitude.  Moreover, this number depends strongly on which field scale is being extracted from the correlation function ($\delta B_\frac{1}{2}$ vs. $\delta B_i$, etc.).

\begin{figure}[t]
\includegraphics{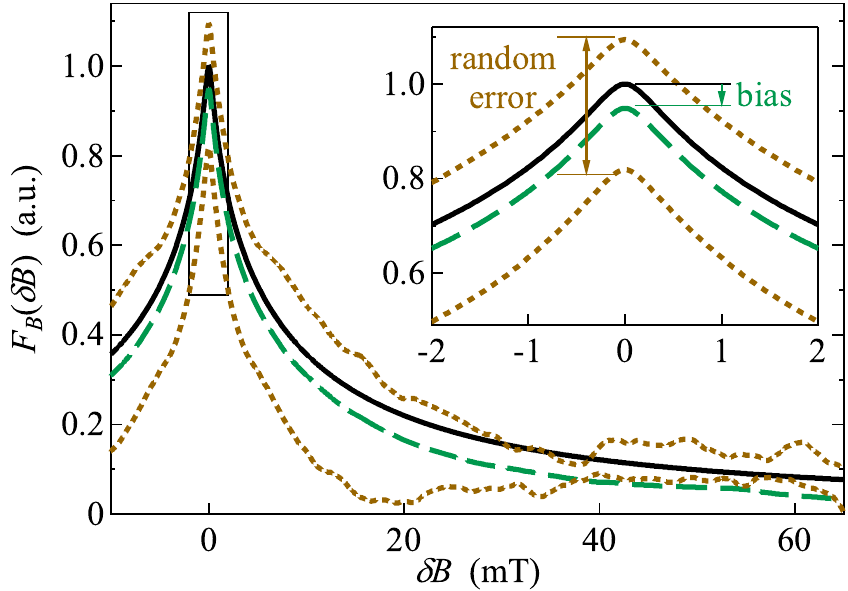}
\caption{\label{fig-errors}Example of statistical errors in UCF analysis for the case $T = 1 ~\mathrm K$, $\tphi = 100 ~\mathrm{ps}$, and $D = 0.03~\mathrm{m^2/s}$.
The {solid line} indicates the UCF correlation function $F_B(\delta B)$. The { dotted lines} show the autocorrelations of two simulated $G(B)$ traces over a $1~\mathrm T$ range in $B$, with mean background subtraction ($\ltotal = \lcut = 1~\mathrm T$; see Appx.~\ref{sec:ucfsim}). The {dashed line} is an average of 100 such autocorrelations ($\ltotal = 100~\mathrm{T}$, $\lcut = 1~\mathrm T$).
{\em Inset}: Expansion of the boxed region, graphically depicting the magnitudes of the two types of error in variance.
}
\end{figure}

\begin{figure}[t]
\includegraphics{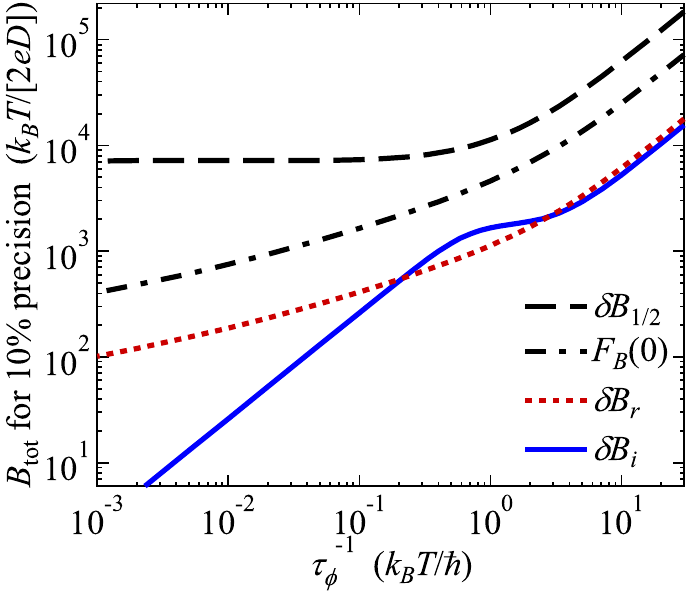}
\caption{\label{fig-error-rand}Guidelines for the required total scan length in field ($\ltotal$) for reaching 10\% standard deviation in the various measures of $F_B(\delta B)$.
As expected, the standard deviation of any measure falls as $\ltotal^{-0.5}$ when $\ltotal$ is increased.
}
\end{figure}

Random errors appear in the correlation function when it is estimated from a finite data set.
Figure~\ref{fig-errors} shows an example of random errors, which appear as fluctuations in the estimated correlation function.
These fluctuations in turn cause uncertainties in all derived parameters such as correlations lengths or variance.
The magnitudes of the random errors depend on the total scanned range of data, $\ltotal$, which may be distributed over multiple independent scans.
As seen in Fig.~\ref{fig-error-rand}, 
the total scan length required for a reliable estimate is different by many orders of magnitude for the various statistical metrics.

\begin{figure}[t]
\includegraphics{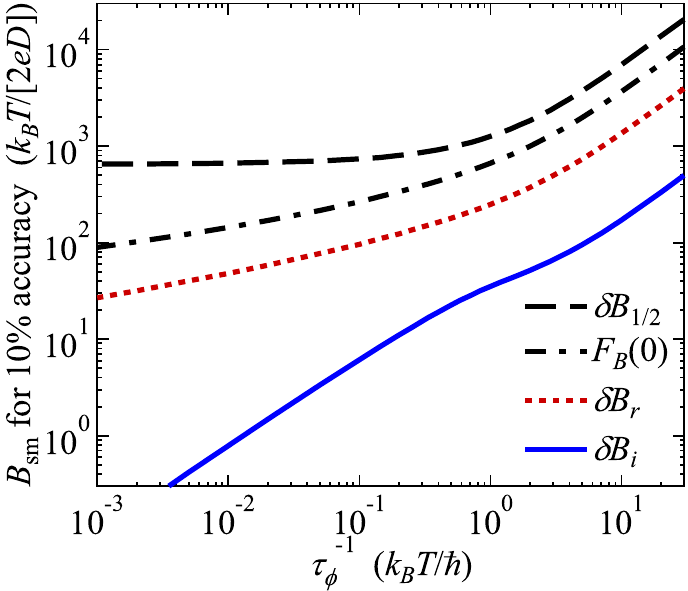}
\caption{\label{fig-error-bias}Guidelines for the required smoothing-length in field ($\lcut$) for reaching $-10\%$ bias in the various measures of $F_B(\delta B)$. Here we have taken mean background subtraction with a scan length $\lcut$.
Biases in $B_\frac{1}{2}$, $B_r$, $F_B(0)$ fall as $\approx \log(\lcut)/\lcut$ when $\lcut$ is increased. Bias in $B_i$ falls as $\lcut^{-2}$.
}
\end{figure}

{ Systematic errors} (biases) occur when the background conductance is estimated from measured conductance data itself by fitting, smoothing, etc.
Background subtraction inevitably affects the correlation function: any smoothed background removes slowly-varying but genuine conductance fluctuations that are longer than a ``smoothing length" $\lcut$, which must be determined from the smoothing algorithm and is necessarily less than $\ltotal$ (see Appx.~\ref{sec:subbg}).
Considering the effects of a smoothed background, the consistent loss of long-ranged fluctuations distorts the analysis in the form of a downward bias (see, e.g., Fig.~\ref{fig-errors}).
This bias directly impacts the accuracy of not only variance ($F_B(0)$) but also $\delta B_\frac{1}{2}$ and $\delta B_r$, whose definitions rely on variance (Fig.~\ref{fig-error-bias}).
$\delta B_i$ is nearly immune to biases since it does not depend on variance.

Figures 3--5 demonstrate the dramatically different sensitivity to errors for half-width and inflection point.
To put this difference in practical terms, consider a typical low temperature UCF measurement of $\tphi$ in a disordered semiconductor, where one might have $T = 1 ~\mathrm{K}$, $\tphi = 100 ~\mathrm{ps}$, and $D = 0.03 ~\mathrm{m^2/s}$. 
To achieve 10\% accuracy (systematic error) using $\delta B_\frac{1}{2}$, an extremely large $\lcut = 3~\mathrm{T}$ would be required even though the value of $\delta B_\frac{1}{2}$ itself is just $5.2~\mathrm{mT}$.  Using $\delta B_i$, on the other hand, 10\% accuracy would be obtained for $\lcut=5~ \mathrm{mT}$, smaller by three orders of magnitude compared to the $\delta B_\frac{1}{2}$ case.
Similarly, 10\% precision in $\tphi$ using $\delta B_i$ would require a total scan length $\ltotal = 300~\mathrm{mT}$, compared to $\ltotal = 45~\mathrm{T}$ for $\delta B_\frac{1}{2}$.


\section{\label{sec:symmetry}Influence of symmetry-breaking disorder}

\begin{figure}[t]
\includegraphics{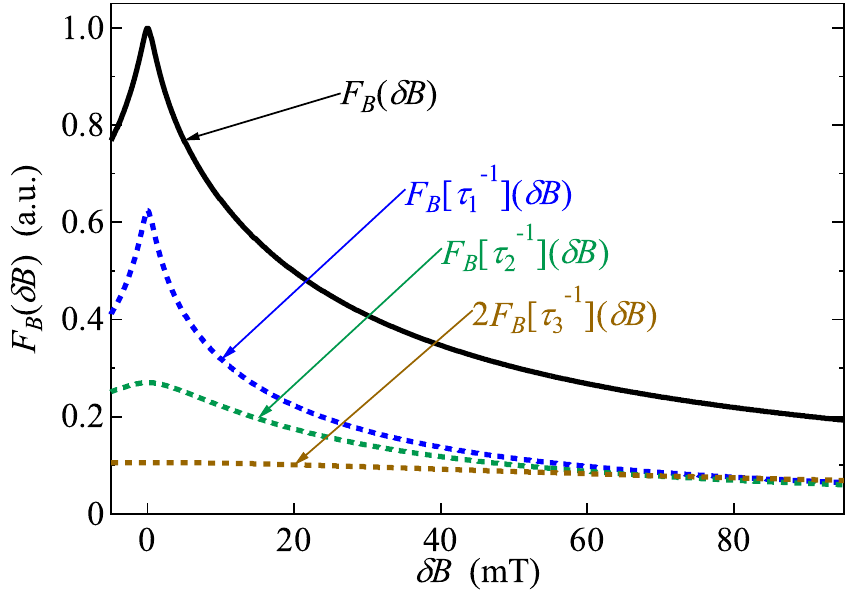}
\caption{\label{fig-corrsym}Contributions to the magnetic field correlation function from three UCF modes at $T=2~\mathrm K$, $D = 0.03~\mathrm{m^2/s}$. The dotted curves show separate modes with $\tau_{1,2,3}^{-1} = \{20~\mathrm{ns^{-1}}, 220~\mathrm{ns^{-1}}, 2120~\mathrm{ns^{-1}} \} $ with $N_{1,2,3} = \{1,1,2\} $, and the solid curve is their sum [Eq.~\eqref{sumsym}].
}
\end{figure}

Many systems of practical interest involve more than one dephasing rate, giving a UCF correlation function that deviates from the behaviour described in Sec.~\ref{sec:single}.
These additional dephasing rates arise from symmetry-breaking static disorders that partially dephase UCF, e.g.,~frozen magnetic impurities,\cite{bobkov} spin-orbit coupling,\cite{chandrasekhar} or valley-orbit coupling.\cite{grapheneucf,grapheneucf2} With such disorder, the correlation function becomes the sum of independent modes with a set of distinct dephasing rates \{$\tau_i^{-1}$\} and degeneracies $\{N_i\}$, in the form
\begin{align}
F(\delta \mu, \delta B) & = N_1 F[\tau_1^{-1}](\delta \mu, \delta B)  \nonumber \\
& \quad{} + N_2 F[\tau_2^{-1}](\delta \mu, \delta B) \nonumber \\
& \quad{} + N_3 F[\tau_3^{-1}](\delta \mu, \delta B) + \cdots .
\label{sumsym}
\end{align}
These modes are often known as the {diffuson singlets} and { diffuson triplets}.\cite{grapheneucf,grapheneucf2,bobkov,chandrasekhar}
With this summation, the field scales of $F$ are determined by a complicated mixture of the temperature and various dephasing rates, so that the considerations of Sec.~\ref{sec:single} may not directly apply. Nevertheless, each independent mode $F[\tau_i]$ is of the type described in Appendix \ref{sec:2dtheory}, so it is straightforward to compute Eq.~\eqref{sumsym} and numerically extract the field scales.

Figure~\ref{fig-corrsym} shows an example involving the three modes appropriate for graphene. We have chosen typical\cite{tikhonenko} values for $T = 2~\mathrm K$: a decoherence rate of $\tau_{\phi}^{-1} = 20~\mathrm{ns^{-1}}$, an intervalley rate of $\tau_{\rm iv} = 100~\mathrm{ns^{-1}}$, and an intravalley rate of $\tau_{*} = 2000~\mathrm{ns^{-1}}$.
The resulting modal dephasing rates are\cite{grapheneucf,grapheneucf2} $\tau_1^{-1} = \tau_{\phi}^{-1}$, $\tau_2^{-1} = \tau_{\phi}^{-1} +2 \tau_{\rm iv}^{-1}$, and $\tau_3^{-1} = \tau_{\phi}^{-1} + \tau_{\rm iv}^{-1} + \tau_{*}^{-1}$. Although the rates are greatly different in magnitude, each mode has a significant contribution to $F_B(\delta B)$ because of thermal smearing (see Appx.~\ref{sec:longrange}).
This causes the variance, half-width, and roundness to differ greatly from the value expected of the dominant mode ($\tau_1$) alone. For instance, $\delta B_\frac{1}{2}$ is twice as large, which would be misinterpreted (by the considerations of Sec.~\ref{sec:single}) as a dephasing rate four times larger than the actual $\tau_1^{-1}$.
The inflection point, however, remains a reliable measure of $\tau_1^{-1}$ even when the additional rates are neglected, with only a 4\% error.

\begin{figure}[t]
\includegraphics{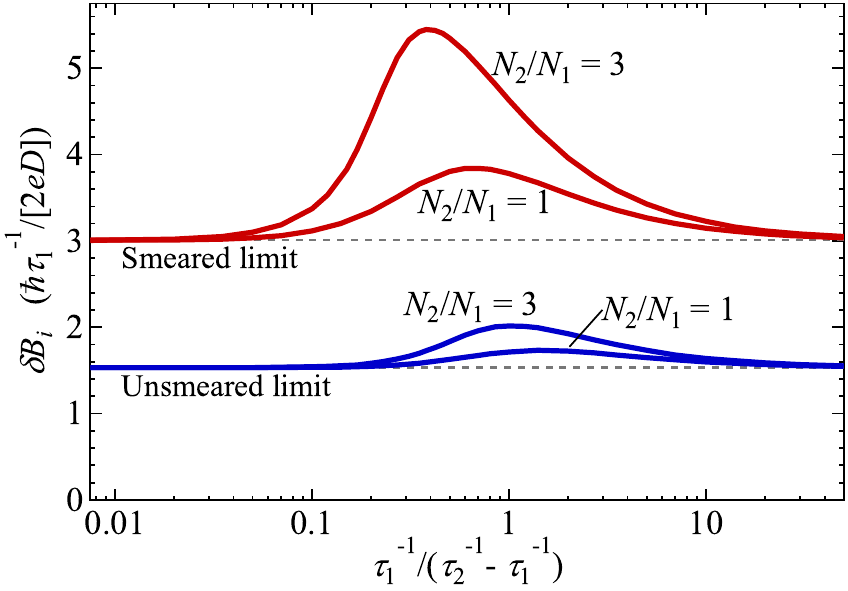}
\caption{\label{fig-combomap}Effect on field correlation's inflection point from the combination of two UCF modes, in smeared ($\kT \geq 10 \hbar \tau_1^{-1}$) and unsmeared ($\kT \leq 0.02 \hbar \tau_1^{-1}$) limits. Dashed lines show the unperturbed inflection point when there is no secondary mode ($N_2 = 0$).
}
\end{figure}

Figure \ref{fig-combomap} examines how $\delta B_i$, computed for the case of only two modes, depends on the relative dephasing rates of the two modes.
Here $\tau_1^{-1}$ might be the decoherence rate from dynamic scatterers that affects all UCF modes, while $(\tau_2^{-1} - \tau_1^{-1})$ could be the extra static symmetry-breaking rate affecting only the second mode. When the symmetry-breaking rate is comparable to the decoherence rate, the inflection point may be  displaced from the value expected of $\tau_1^{-1}$ alone. The degree of displacement, however, never exceeds a factor of 1.8 (this only occurs if $N_2/N_1 = 3$ and $\hbar\tau_1^{-1} \ll \kT$).
Fig.~\ref{fig-combomap} also gives a simple rule for $\delta B_i$: the secondary mode can be neglected when it dephases much more rapidly than the primary mode, $ \tau_2^{-1} \gtrsim 10\tau_1^{-1}$; such a simple rule does not apply for other aspects, e.g.,~$F_B(0)$ or $\delta B_\frac{1}{2}$. In the opposite limit, when dephasing rates for all modes are similar, $\tau_2^{-1} \approx \tau_1^{-1}$, the field scales are determined by the considerations of Sec.~\ref{sec:single} with the single dephasing rate $\tau_1^{-1}$ (or $\tau_2^{-1}$), and $F_B(0)$ is only increased by a trivial factor.\cite{grapheneucf,grapheneucf2}

The statistical errors (Sec.~\ref{sec:errors}) are also modified by the presence of symmetry-breaking disorder, often becoming much larger. The considerations of Appx.~\ref{sec:errorstheory} may be used to evaluate errors in a general correlation function with multiple dephasing rates.

\section{\label{sec:1d}Comparison to quasi-1D case}

The characteristics of quasi-2D UCF can be compared to the quasi-1D regime, which occurs when the material is shaped as a long and very narrow strip with a width $W$ that is much smaller than the dephasing length $L_\phi = \sqrt{D \tphi}$, yet where the length between contacts is longer than $L_\phi$.
An advantage of quasi-1D UCF (especially useful in metals) is that shaping the material into a wire produces a lower background conductance which allows UCF to appear with higher contrast.
The quasi-1D correlation function $F_B(\delta B)$, which is known exactly in both the smeared and unsmeared limits, demonstrates essentially single-scale behaviour, falling as $1/|\delta B|^3$ at high $\delta B$.\cite{beenakker1d} This explains why the half-width performs well as a measure of coherence in the quasi-1D system: its statistical errors are only somewhat higher than the inflection point for a given range of field. 

For completeness, we note the values of inflection point for the quasi-1D system in the `dirty' regime, where the elastic mean free path is smaller than $W$. The quasi-1D energy correlation function in Ref.~\onlinecite{falkoucf} gives the inflection point $\delta \mu_i = 0.549 \hbar/\tphi$ in the unsmeared limit; this converges on the universal value $\delta \mu_i = 2.14 \kT$ in the smeared limit. As for the the magnetic field inflection point,
the formulas in Ref.~\onlinecite{beenakker1d} yield $\delta B_i = \sqrt{3}\hbar/(e W L_\phi)$ in the unsmeared limit, and $\delta B_i = \sqrt{6}\hbar/(e W L_\phi)$ in the smeared limit.


\section{Conclusion}

In closing, the analysis of UCF is a valuable tool for experimental studies of dephasing, as it can provide information on both the electron temperature and the dephasing time under a wide range of conditions. We have shown that the correlation function inflection points in energy and magnetic field provide an accurate and efficient way to do so. This inflection point method will be especially useful in the experimental study of the magnetic field dependence of coherence in quasi-2D systems.

\appendix

\section{\label{sec:2dtheory}Theoretical quasi-2D UCF correlation}

The procedure for calculating the theoretical $F$ was originally formulated by Stone, Lee, and Fukuyama,\cite{lsf} and first computed for the quasi-2D case by Bergmann.\cite{bergmann} We present a  different numerical approach than that in Ref.~\onlinecite{bergmann}, by first calculating the unsmeared correlation function analytically.
Of special interest is the asymptotic behaviour in the smeared limit, discussed in subsection \ref{sec:longrange}.

The main result is \eqref{ucfcorr2Danalytical}, which when combined with the thermal smearing convolution \eqref{ucfcorrsmeardimless} provides the correlation function of UCF \eqref{ucfcorrsmear}. The correlation function here is computed for the case of a single dephasing rate, and may be used as a building block in more complicated situations of multiple dephasing rates (Sec.~\ref{sec:symmetry}) or other modifications (e.g.,~Zeeman splitting).

\subsection{Exact unsmeared correlation function}
Consider a rectangular conductor with length $L_x$ between the source and drain contacts (at opposing edges) and width $L_y$. We begin by analyzing the fluctuations in the quantity $G_0(E,B)$, which is the source-drain conductance at a fixed energy $E$ (with thermal smearing not yet taken into account) and in magnetic field $B\hat z$.
The ensemble-averaged correlations in the conductance fluctuations, $\delta G_0 = G_0 - \overline{G_0}$, are given by\cite{mpep}
\begin{align}
F_0(\delta E, \delta B) &\equiv \overline{\delta G_0(E,B)\, \delta G_0(E+\delta E, B + \delta B)} \nonumber\\
& = C\frac{e^4}{h^2} \frac{4D^2 }{ L_x^4} \sum_{n} \left[\frac{1}{|\lambda_n|^2} + \frac{1}{2} {\rm Re}\frac{1}{\lambda_n^2}\right],
\label{ucfcorrstone}
\end{align}
where $C$ is a constant prefactor depending on the number of intact symmetries in the system.\footnote{$C=1$ if there are no spin/valley/time-reversal symmetries.}
The $\lambda_n$ are the eigenvalues of the diffusion equation
\begin{equation}
\left[  D \left(i \vec \nabla - \frac{e}{\hbar} \vec  A(\vec r) \right)^2 + \tphii - i \frac{\delta E}{\hbar}  \right] Q(\vec r)
 = \lambda_n Q(\vec r),
\label{ucfeigenequation}
\end{equation}
where $\vec\nabla \times \vec A = \delta B \hat z $, and with Dirichlet and Neumann boundary conditions at the contact and vacuum edges, respectively.

The primary assumption of the quasi-2D case is that we can ignore the boundary conditions, in which case the eigenfunctions $Q(\vec r)$ of \eqref{ucfeigenequation} are Landau levels.\cite{lsf,bergmann,mpep} The ``cyclotron frequency'' here is $2De|\delta B|/h$, giving the series of eigenvalues
\begin{equation}
\lambda_n = \frac{2De|\delta B|}{\hbar}\Big( n+\frac{1}{2}\Big) + \tphii - i\frac{\delta E}{\hbar},
\end{equation}
for non-negative integers $n$, with the degeneracy of $e|\delta B|L_x L_y/h$ for each level.
This places $L_x$ and $L_y$ dependence only in the prefactor of $F_0$, allowing us to write
\begin{equation*}
F_0(\delta E, \delta B) = C \frac{e^4}{h^2} \frac{ L_y D \tphi}{ L_x^3} \mathfrak{F}_0(\varepsilon,\beta).
\label{ucfcorr2Ddimless}
\end{equation*}
where we have introduced a scale-independent function $\mathfrak{F}_0(\varepsilon, \beta)$ in terms of a dimensionless energy and field,
\begin{equation*}
\varepsilon \equiv \delta E \cdot \tphi/\hbar, \qquad \beta \equiv |\delta B| \cdot 2 e D \tphi/\hbar.
\label{epsilonbeta}
\end{equation*}

The sum in \eqref{ucfcorrstone} may be solved exactly in terms of the complex digamma function $\digammaa(z)$ and its derivative, $\trigamma(z)$. For $\mathfrak{F}_0$ this solution is written as
\begin{align}
\mathfrak{F}_0(\varepsilon,\beta)
& =  \frac{1}{\pi\varepsilon}{\rm Im\!}\left[
    \digammaa\bigg(\frac{1}{2} + \frac{1+i\varepsilon}{\beta}\bigg)\right] \nonumber\\
& \quad + \frac{1}{2\pi\beta} {\rm Re\!} \left[
    \trigamma\bigg(\frac{1}{2} + \frac{1+i\varepsilon}{\beta}\bigg)\right].
\label{ucfcorr2Danalytical}
\end{align}
This expression does not evaluate when either $\beta$ or $\varepsilon$ are zero; taking limits, one obtains the variance $\mathfrak{F}_0(0,0) = \frac{3}{2\pi}$ and one-parameter correlations\cite{bergmann,mpep}
\begin{eqnarray}
\label{ucfcorr2Dlimite}
\mathfrak{F}_0(\varepsilon,0) &= &\frac{\tan^{-1} \varepsilon}{\pi\varepsilon} + \frac{1}{2\pi}(1+\varepsilon^2)^{-1},\\
\quad \mathfrak{F}_0(0,\beta)& =& \frac{3}{2\pi\beta} \trigamma\Big(\frac{1}{2} + \frac{1}{\beta}\Big).
\label{ucfcorr2Dlimitb}
\end{eqnarray}

\subsection{Thermal smearing effect}

The direct effect of temperature is to average the bare conductance $G_0$ over a range of values $\delta E \sim \delta \mu \pm \kT$.
In detail, the measured conductance $G$ is determined by the thermal smearing convolution $$G(\mu,B) = \int_{-\infty}^{\infty} \!\!{\rm d}E\, f_F'(E-\mu) G_0(E, B),$$
where
$f_F'(\delta E) = \frac{1}{4\kT}\sech^2(\frac{1}{2\kT}\delta E)$ is the Fermi function.
The smeared correlation function, $F(\delta \mu,\delta B)$, is then obtained by a convolution of $F_0(\delta E,\delta B)$ with the function
$\frac{1}{\kT}\kappa(\frac{\delta E}{\kT})$
where $\kappa(x) = \frac{1}{2}(\frac{x}{2}\coth \frac{x}{2} - 1)/\sinh^2 \frac{x}{2}$.\cite{mpep}

In terms of scale-independent variables, the thermal smearing effect modifies $\mathfrak{F}_0 \rightarrow \mathfrak{F}_{\mathfrak{T}}$ by the convolution:
\begin{equation}
\mathfrak{F}_{\mathfrak{T}}(\varepsilon',\beta) = \int_{-\infty}^{\infty} \!\!{\rm d}\varepsilon\, \frac{\kappa(\varepsilon/\mathfrak{T}) }{\mathfrak{T}}   \mathfrak{F}_0(\varepsilon-\varepsilon',\beta).
\label{ucfcorrsmeardimless}
\end{equation}
where we have defined a dimensionless temperature
$$\mathfrak{T} \equiv \kT \cdot \tphi/\hbar. $$
Equation \eqref{ucfcorrsmeardimless} is quickly computed by way of fast fourier transforms and the convolution theorem.\footnote{The transform $\int_{-\infty}^{\infty} \!\mathrm d\varepsilon\, \frac{1}{a}\kappa(\varepsilon/a) e^{i \varepsilon t} = [\pi a t /\sinh(\pi a t)]^2$ allows thermal smearing to be easily applied in fourier space.}
The scaled result of thermal smearing is then written as
\begin{equation}
F(\delta \mu, \delta B) = C\frac{e^4}{h^2} \frac{L_y D \tphi}{L_x^3} \, \mathfrak{F}_{\mathfrak{T}} ( \varepsilon',\beta ).
\label{ucfcorrsmear}
\end{equation}
Note that here, $\varepsilon' = \delta \mu \tphi/\hbar$  (not $\delta E$).

\subsection{\label{sec:longrange}Highly smeared behaviour}


We now examine the behaviour of quasi-2D UCF under a large amount of thermal smearing, $\mathfrak T \gg 1$. As seen in Eq.~\eqref{ucfcorr2Dlimite}, $\mathfrak{F}_0(\varepsilon,0) \approx |2\varepsilon|^{-1}$ for large $\varepsilon$, and this plays a critical role when the convolution \eqref{ucfcorrsmeardimless} is applied. The $1/|\varepsilon|$ behaviour is responsible for the long-ranged and multi-scale nature of $F_B(\delta B)$ (Sec.~\ref{sec:single}), its unusually large statistical errors (Sec.~\ref{sec:errors}), and the high sensitivity to broken-symmetry modes (Sec.~\ref{sec:symmetry}).

An immediate consequence of the $1/|\varepsilon|$ behaviour is the logarithmic form of the variance under thermal smearing.\cite{bergmann,mpep} 
The asymptotic behaviour of variance (from Eq.~\eqref{ucfcorrsmeardimless}) in the smeared limit is
\begin{equation}
\mathfrak{F}_{\mathfrak{T}}(0,0) \approx \frac{1}{6\mathfrak{T}} \ln(C_0\mathfrak{T}), \quad \mathfrak{T} \gg 1
\label{varianceasymptotic}
\end{equation}
where we find numerically that $C_0 = 4.1$. This gives an extremely weak $\tphi$-dependence of the measured variance, $F(0,0) \propto \ln(T\tphi)/T$.

More generally, the correlation function \eqref{ucfcorr2Danalytical} falls as $\mathfrak{F}_0(\varepsilon,\beta) \sim \textrm{min}\{|\varepsilon|^{-1}, \beta^{-1}\}$ when either argument is large. In this case we can approximate \eqref{ucfcorr2Danalytical} by taking $\frac{1}{2}+(1+i\varepsilon)/\beta \approx \frac{1}{2}+ i\varepsilon/\beta$, giving a $\tphi$-independent form:
\begin{equation*}
\mathfrak{F}_0(\varepsilon,\beta) \approx \frac{\pi}{2\beta} \mathfrak{f}(\pi\varepsilon/\beta),\quad \beta \gg 1~\mathrm{or}~\varepsilon \gg 1
\label{f0highfield}
\end{equation*}
where Euler's reflection formula gives $\mathfrak{f}(x) = \frac{1}{x}\tanh x + \tfrac{1}{2}\sech^2 x$.
When a high amount of thermal smearing is applied, the first term in $\mathfrak{f}(x)$ dominates. This gives the intermediate-field behaviour of the field correlation function in the smeared limit,
\begin{equation}
\mathfrak{F}_{\mathfrak{T}}(0,\beta) \approx \frac{1}{6\mathfrak{T}} \ln(C_1\mathfrak{T}/\beta), \quad  \mathfrak{T} \gg \beta \gg 1
\label{fthighfield}
\end{equation}
where we find numerically that $C_1 = 29.1$. From \eqref{varianceasymptotic} and \eqref{fthighfield} one obtains the asymptotic half-width listed in Table~\ref{table-asymptotics}, $\beta_\frac{1}{2} \approx C_1(\mathfrak{T}/C_0)^\frac{1}{2}$.

The correlation function derivative $\frac{\partial}{\partial \beta}\mathfrak{F}_0(\varepsilon,\beta)$ approaches zero rapidly, as $ \varepsilon \sech^2(\pi\varepsilon/\beta)$, for large $\varepsilon$. As a result the thermal smearing convolution imposes a simple behaviour for this field derivative:
\begin{equation}
\frac{\partial}{\partial \beta}\mathfrak{F}_{\mathfrak{T}}(\varepsilon,\beta) \approx \mathfrak{h(\beta)} \frac{\kappa(\varepsilon/\mathfrak{T})}{\mathfrak{T}}, \quad \mathfrak{T} \gg \beta~\mathrm{and}~\mathfrak{T} \gg 1
\end{equation}
for a function $\mathfrak h(\beta) = \int_{-\infty}^{\infty} \!\mathrm d\varepsilon\, \frac{\partial}{\partial \beta}\mathfrak{F}_0(\varepsilon,\beta)$. This separable $\beta$ dependence explains why the inflection point $\beta_i$ is constant for large $\mathfrak T$. The zero-field curvature can also be computed, $\frac{\mathrm d^2}{\mathrm d \beta^2}\mathfrak{F}_{\mathfrak{T}}(0,\beta)|_{\beta = 0} \approx \mathfrak{h}'(0)/(6\mathfrak T) $, which provides the smeared-limit asymptotic roundness when combined with Eq.~\eqref{varianceasymptotic}, $\beta_r \approx [2\ln(C_0\mathfrak{T})/\mathfrak{h}'(0)]^{0.5}$. Numerically we find $\mathfrak{h}'(0) = 1/12$.

\section{\label{sec:errorstheory}Theory of statistical errors in measurements of correlation functions}

This appendix explores the statistical errors that occur in the estimation of the correlation function from a generalized fluctuating quantity $G(x)$. We will use upper-case $G$ and $F$ to represent the error-free quantity and its ideal correlation, and lower-case $g$ and $f$ to represent estimated values.
The overline notation ($\overline{\delta G\, \delta G}$, $\overline{f}$, $\overline{x_i}$ etc.) in this section refers specifically to an {\em ensemble} average (average over disorder configurations). The ergodicity assumption, which equates this with an average over $x$, does not apply in subsection \ref{sec:subbg} where we are essentially calculating the inaccuracy of the ergodicity assumption under weak violations.

A typical experiment measures conductance $G(x)$ over a limited range $x = 0 \cdots L$. Here $x$ is a parameter such as $\mu$ or $B$. Next, a background estimate $g_B(x)$ is computed from $G(x)$, then subtracted to yield the estimated fluctuations, $\delta g(x) = G(x)-g_B(x)$.
Finally, the correlation function is estimated as
$$f(\delta x) = \frac{1}{L-\delta x} \int_0^{L-\delta x}\!\!\mathrm dx\, \delta g(x) \delta g(x+\delta x),$$
and the correlation length estimators $x_\frac{1}{2}$, $x_r$ and $x_i$ are extracted from $f(\delta x)$.
This estimator $f(\delta x)$ differs from the true correlator $F(\delta x) = \overline{\delta G(x)\,\delta G(x+\delta x)}$ for two statistical reasons.
First, $\delta g(x)$ has lost some information from the true fluctuations $\delta G(x)$ due to the background subtraction, which leads to systematic error in $f(\delta x)$. Second, the limited range $L$ leads to random errors in $f(\delta x)$ depending on the particular realization of $\delta G(x)$. These two error mechanisms will be addressed separately.

\subsection{\label{sec:subbg}Background subtraction errors}

To calculate the background subtraction bias, we consider the simplest possible procedure which is to subtract the mean of $G(x)$ over the measured interval $L$. The results of this section can be extrapolated to a general background fitting procedure by taking an effective $L \approx L_{\rm sm}$. For example, if a polynomial fit to $G(x)$ is subtracted, then $L_{\rm sm} = L_{\rm total}/n$, where $L_{\rm total}$ is the measured range and $n$ is the number of degrees of freedom in $g_B(x)$ [e.g.,~$n=3$ for a parabolic $g_B(x)$].

The mean background is $g_B(x) = \frac{1}{L} \int_0^{L}\mathrm dx'\, G(x')$, which gives an error $[\delta g(x) - \delta G(x)] =  - \frac{1}{L} \int_0^{L}\mathrm dx'\, \delta G(x')$ in the estimated fluctuations.
The resulting systematic error in the autocorrelation function $f(\delta x)$ is approximately constant for $\delta x \ll L$: To first order,
\begin{equation}
\mathrm{Bias}\{f(\delta x)\} = \overline{f(\delta x)} - F(\delta x) \approx -  F(0) \frac{ x_L }{L}.
\label{biasf}
\end{equation}
Here, $x_L$ is a characteristic correlation length, defined basically by the area under the correlation function:
\begin{equation}
x_L = \int_{-L}^{L} \!\!\mathrm dz\, \Big(1-\frac{|z|}{L}\Big) \frac{F(z)}{F(0)}
\label{biasx}
\end{equation}

For short-ranged correlation functions, $x_L$ would be a constant for large $L$, and so the systematic error \eqref{biasf} would fall as $1/L$; this bias then would be similar to the well-known sample variance bias from independent sample statistics, agreeing with the intuition of $G(x)$ containing ``many independent fluctuations'', each having length $x_L$.
In the quasi-2D UCF case, however, $F(\delta x)$ only falls as $1/\delta x$ [see \eqref{ucfcorr2Dlimite}, \eqref{ucfcorr2Dlimitb}] so the value of $x_L$ diverges logarithmically as $L$ increases. Hence the similarity with independent sample statistics does not hold for the quasi-2D UCF variance bias, as there is no well-defined ``independence length''.

The bias in variance $f(0)$ leads to direct effects on the half-width estimator ($x_\frac{1}{2}$) and the roundness estimator ($x_r$), as these are both sensitive to the absolute variance.
The roundness estimator $x_r$ is biased by $\mathrm{Bias}\{ x_r\} \approx - \frac{1}{2} x_r x_L/L $.
The bias in $x_\frac{1}{2}$ is given by
$$\mathrm{Bias}\{ x_\frac{1}{2}\}  \approx \frac{F(0)}{2F'(x_\frac{1}{2})} \frac{ x_L}{L}. $$
The inflection point estimator $x_i$ depends only on the second derivative of $f(\delta x)$, so to first order $x_i$ has no bias; taking into account higher order terms omitted from \eqref{biasf} we obtain $\mathrm{Bias}\{ x_i\} \approx -(2/L^2) F(x_i)/F'''(x_i)$.

\subsection{Random errors}

Next we suppose the background has been determined perfectly, giving us the exact fluctuations: $\delta g(x) = \delta G(x)$.
Although we have $\overline{ f(\delta x)} = F(\delta x)$ in this case, the measured $f(\delta x)$ will have random deviations from $F(\delta x)$ due to the limited data set.
The random fluctuations in $f(\delta x)$ can be expressed in terms of a two-point correlator $\overline{f(\delta x_1)f(\delta x_2)}$. If $\delta G(x)$ is gaussian (as is the case for UCF\cite{mpep}) then Isserlis' theorem yields
\begin{align}
& \overline{f(\delta x_1)f(\delta x_2)} - \overline{f(\delta x_1)}\: \overline{f(\delta x_2)} \nonumber \\
&\qquad{} = \frac{1}{L}  [H(\delta x_1-\delta x_2) + H(\delta x_1+\delta x_2)],
\label{twopointcorrelator}
\end{align}
for a dataset of large length $L \gg \delta x_1, \delta x_2$,
where $H(\delta\delta x)$ is a higher-order correlator, defined as 
\begin{equation}
H(\delta\delta x) = \int_{-\infty}^{\infty} \!\!\mathrm dz\, F(z) F(z+\delta\delta x).
\end{equation}
Equation \eqref{twopointcorrelator} and its derivatives allow the determination of random errors in any aspect of $f(\delta x)$, including its correlation lengths. For instance, the random error in variance [$f(0)$] is given by $\mathrm{Var}\{f(0)\} = 2H(0)/L$.

The half-width estimator $x_\frac{1}{2}$ is sensitive to the errors in both $f(\overline{ x_\frac{1}{2}})$ and $f(0)$, modulated by the local slope $F'(\overline{ x_\frac{1}{2}})$, giving
\begin{align}
\mathrm{Var}\{x_\frac{1}{2}\}
&= \mathrm{Var}\bigg\{\frac{f(\overline{ x_\frac{1}{2}}) - \tfrac{1}{2}f(0)}{F'(\overline{ x_\frac{1}{2}})}\bigg\} \nonumber \\
&= \frac{1}{L} \frac{\tfrac{3}{2}H(0)+H(2\overline{ x_\frac{1}{2}})-2H(\overline{ x_\frac{1}{2}})}{F'(\overline{ x_\frac{1}{2}})^2}.
\end{align}

The error in the inflection point estimator $x_i$ depends on the fluctuation of $f''(\overline{x_i})$, modulated by the local slope $F'''(\overline{x_i})$.
\begin{align}
\mathrm{Var}\{x_i\}
& =  \mathrm{Var}\{f''(\overline{x_i})/F'''(\overline{x_i})\} \nonumber \\
& = \frac{1}{L} \frac{H''''(0) + H''''(2 \overline{x_i})}{F'''(\overline{x_i})^2}.
\end{align}

The roundness estimator, $x_r = \sqrt{2f(0)/|f''(0)|}$, is influenced by changes in both $f(0)$ and $f''(0)$:
\begin{align}
\mathrm{Var}\{x_r\}
&= \mathrm{Var}\bigg\{\frac{f(0)}{\sqrt{2F(0)|F''(0)|}} + \frac{\sqrt{F(0)}f''(0)}{\sqrt{2|F''(0)|^3}} \bigg\} \nonumber \\
&= \frac{1}{L}\left[\frac{H(0)}{F(0)|F''(0)|} + \frac{F(0)H''''(0)}{|F''(0)|^3} + \frac{2H''(0)}{ F''(0)^2 } \right].
\end{align}

\subsection{\label{sec:ucfsim}Random UCF simulation}

This subsection describes a method to generate UCF traces $\delta G(x)$ from a given correlation function $F(\delta x)$, useful in error analyses such as Fig.~\ref{fig-errors}.

First, $F(\delta x)$ is computed over a very wide range $-L\cdots L$. Next, the discrete fourier transform $\mathrm{FFT}_n\{F(\delta x)\}$ is computed, which is the power spectrum of UCF (by the Wiener-–Khinchin theorem).
Fourier amplitudes of UCF, $G_n = (a_n + i b_n) \sqrt{\mathrm{FFT}_n\{F(\delta x)\}} $, are then generated with random Gaussian numbers $a_n$, $b_n$ with $\overline{a_n^2} = \overline{b_n^2} = \frac{1}{2}$.
Finally, the conductance fluctuations are given by the inverse transform $\delta G(x) = \mathrm{IFFT}(x)\{G_n\}$.



\end{document}